\documentclass[prl,twocolumn,aps]{revtex4-1}
\newcommand{\pa}{\partial}

\usepackage{bm,graphicx,color,amsmath}

\begin{document}

\title{Instabilities at Frictional Interfaces: Creep Patches, Nucleation and Rupture Fronts}

\author{Yohai Bar-Sinai$^1$, Robert Spatschek$^2$, Efim A. Brener$^{1,3}$ and Eran Bouchbinder$^1$}
\affiliation{$^1$Chemical Physics Department, Weizmann Institute of Science, Rehovot 76100, Israel
\\$^2$Max-Planck-Institut f\"ur Eisenforschung GmbH, D-40237 D\"usseldorf, Germany
\\$^3$Peter Gr\"unberg Institut, Forschungszentrum J\"ulich, D-52425 J\"ulich, Germany}


\begin{abstract}
The strength and stability of frictional interfaces, ranging from tribological systems to earthquake faults, are intimately related to the underlying spatially-extended dynamics. Here we provide a comprehensive theoretical account, both analytic and numeric, of spatiotemporal interfacial dynamics in a realistic rate-and-state friction model, featuring both velocity-weakening and velocity-strengthening behaviors. Slowly extending, loading-rate dependent, creep patches undergo a linear instability at a critical nucleation size, which is nearly independent of interfacial history, initial stress conditions and velocity-strengthening friction. Nonlinear propagating rupture fronts -- the outcome of instability -- depend sensitively on the stress state and velocity-strengthening friction. Rupture fronts span a wide range of propagation velocities and are related to steady state fronts solutions.
\end{abstract}


\maketitle

\textit{Introduction --- } Predicting the strength and stability of frictional interfaces is an outstanding problem, relevant to a broad range of fields -- from biology and nano-mechanics to geophysics.
Recent modeling efforts \cite{Marone1998, Lapusta2000, Persson2000a, Cocco2002, Lapusta2003, Rubin2005, Baumberger2006, Ampuero2008, Ben-Zion2008, Braun2009, Shi2010, Bizzarri2010, Tromborg2011a, Amundsen2011, Lorenz2012, Capozza2012, Kammer2012, Otsuki2013, Capozza2013, Vanossi2013} and novel laboratory experiments \cite{Berthoud1999, Ohnaka2000, Rubinstein2004, Nakatani2006, Rubinstein2007, Reches2010, Popov2010, Nielsen2010, Ben-David2010a, Ben-David2010b, Ben-David2011a,Nagata2012, Chang2012} have revealed complex spatiotemporal dynamics that precede and accompany interfacial failure. In particular, frictional instabilities that mark the transition from creep-like motion to rapid slip and a variety of emerging rupture fronts have been observed. Quantitatively understanding these complex dynamics and their dependence on geometry, external forcing, system's history and constitutive behavior of the frictional interface remains an important challenge.

In this Letter we present a theoretical analysis of a simple, yet realistic, quasi-1D rate-and-state model \cite{Dieterich1978, Ruina1983} in which friction is velocity-weakening at low slip velocities and crosses over to velocity-strengthening at higher velocities \cite{Weeks1993,Heslot1994,BarSinai2012}. Using combined analytic and numeric tools we elucidate the physics of a sequence of instabilities at a frictional interface. In particular, we study the dynamics of slowly extending creep patches \cite{Bakun1980,Dieterich1986,Cao1986,Sammis2001,Kohli2011}, their stability, and the emerging nonlinearly propagating rupture fronts.

\textit{The model --- } The friction model we study is the realistic rate-and-state model introduced in \cite{BarSinai2012}, which is briefly presented here. The spatially-extended interface between two dry macroscopic bodies is composed of an ensemble of contact asperities whose total area $A_r$ is much smaller than the nominal contact area $A_n$ \cite{Bowden2001}. The normalized real contact area, $A\!\equiv\!A_r/A_n\!\ll\!1$, is given as $A(\phi)\!=\!\left[1\!+\!b\log\left(1\!+\!\phi/\phi^*\right)\right]\sigma/\sigma_H$, where $\phi$ is a state variable quantifying the typical time passed since the contact was formed (i.e. its ``age''). $\sigma$ is the normal stress, $\sigma_H$ is the hardness, $b$ is a dimensionless material parameter and $\phi^*$ is a short time cut-off \cite{Nakatani2006,Ben-David2010b}. The frictional resistance stress $\tau$ is decomposed as $\tau\!=\!\tau^{el}\!+\!\tau^{vis}$, where $\tau^{el}$ is related to elastic deformation of the contact asperities and $\tau^{vis}$ to their rheological response. The latter is related to thermally-activated processes and is given by $\tau^{vis}(v,\phi)\!=\!\eta\,v^* A(\phi)\log\left(1\!+\!v/v^*\right)$ \cite{Baumberger1999, Rice2001, Baumberger2006}, where $v$ is the slip velocity, $\eta$ is a viscous-friction coefficient and $v^*$ is a small velocity scale.
\begin{figure*}
 \includegraphics[width=\textwidth]{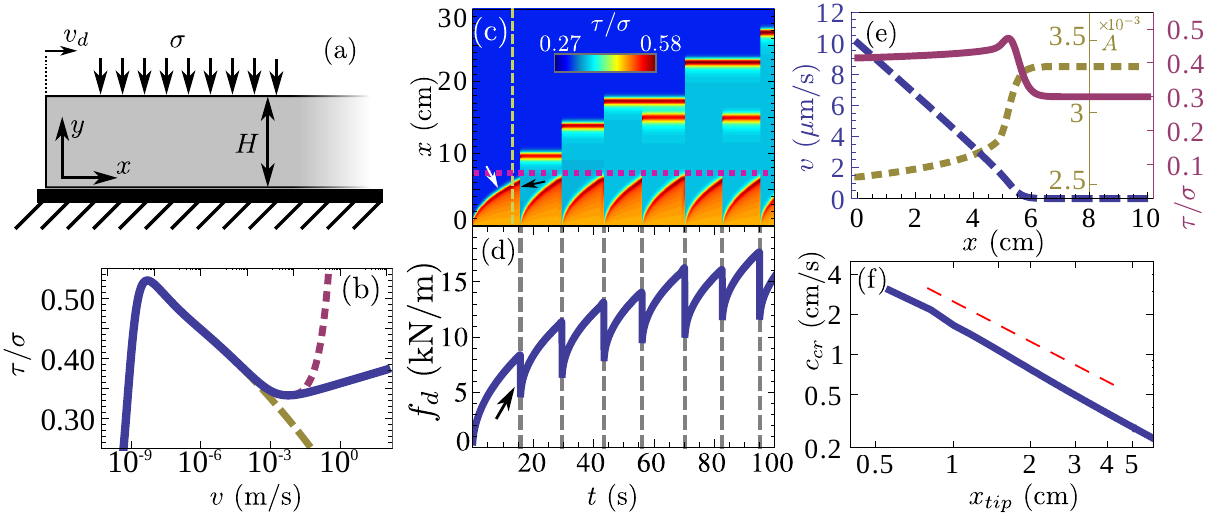}
 \caption{\textbf{General phenomenology and creep patches.} (a) The geometry and loading configuration studied here. (b) Steady-state sliding friction vs. slip velocity $v$ (solid blue line). The dashed yellow line shows steady friction which is purely velocity-weakening. The dotted purple line shows stronger-than-logarithmic (linear) strengthening. See text for details. (c) Spatiotemporal evolution of $\tau(x,t)/\sigma$. The blue regions correspond to the background stress $\tau\!=\!300$kPa. (d) The tangential force per unit width $f_d(t)$. (e) Example of $v(x,t)$, $\tau(x,t)/\sigma$ and $A(x,t)$ of the creep patch at the time marked by a vertical dashed line in panel (c)). (f) $c_{cr}$ vs. $x_{tip}$ in log scale. The dashed line shows a slope of -1.}
\label{fig:fig1}
\end{figure*}

The dynamic evolution equations for the friction variables take the form \cite{Ruina1983, BarSinai2012,Baumberger2006}
\begin{eqnarray}
\!\!\dot\phi \!=\!1\!-\!|v|\frac{\phi}{D}\,g(\tau, v),~~~\dot\tau^{el}\!=\!\frac{\mu_0}{h} A(\phi) v \!-\! |v| \frac{\tau^{el}}{D} \,g(\tau, v). \label{eq:tauel}
\end{eqnarray}
Here $D$ is a characteristic slip distance, $\mu_0$ is the interfacial elastic modulus and $h$ is the effective height of the interface. To understand the role of $g(\tau, v)$, first set it to zero. Then, the equations can be readily integrated to yield $\phi\!=\!t$, which corresponds to the well-established logarithmic aging of $A(t)$, and an elasticity relation $\tau^{el}\!\simeq\! \mu_0 A(t) u/h$ (where $A(t)$ varies much slower than the elastic response and $u$ is the slip displacement. Recall that $\dot{u}\!=\!v$). These relations describe the response of the interface in the absence of irreversible slip. When $g(\tau, v)\!=\!1$, the second terms on the right-hand-side of Eqs. (\ref{eq:tauel}) describe the breakage of contact asperities accompanied by irreversible slip over a length $D$ on a time scale $D/v$. Therefore, $g(\tau, v)$ plays the role of an effective threshold for the onset of irreversible slip. In \cite{BarSinai2012}, $g(\tau, v)$ described a sharp threshold in terms of
the stress $\tau$. Here, we choose $g(\tau, v)\!\equiv\!\sqrt{1\!+\!v_0^2/v^2}$, with an extremely small $v_0\!=\!10^{-9}$m/s. Thus, $|v|\,g(\tau, v)$ changes from $v_0$ for $v\!\to\!0$ to $|v|$ for $v\!\gg\!v_0$. Our results are insensitive to this choice of $g(\tau, v)$.

Consider a rigid substrate and a long elastic body (in the x-direction) of height $H$ (in the $y$-direction) pressed against it by a constant normal stress $\sigma$ applied at $y\!=\!H$, see Fig. \ref{fig:fig1}a. The friction law formulated above describes the interface at $y\!=\!0$. The elastic body is described by Hooke's law and its force balance equation, in the limit of small $H$, reads
\begin{equation}
\rho H \ddot u =\bar{G}(\nu) H\partial_{xx}u-\tau\label{eq:udot}\ ,
\end{equation}
where $\rho$ is the mass density and $\bar{G}(\nu)$ is an effective elastic modulus depending on Poisson's ratio $\nu$ and proportional to the shear modulus $G$ \cite{supp}. In this quasi-1D approximation, $\sigma$ is space- and time-independent.

The material parameters we use below were extracted from extensive experimental data of PMMA. The typical slip distance is set to $D\!=\!0.5\mu$m and the rest of the parameters appear in \cite{supp}. The steady sliding friction curve (obtained by setting to zero the time derivatives in Eqs. (\ref{eq:tauel})) is shown in Fig. \ref{fig:fig1}b (solid line). The curve has a peak at extremely small slip velocities (related to $v_0$), which we believe to be a generic feature of friction \cite{Estrin1996, Shimamoto1986}, though it is of no significance here [45,46]. Moreover, the curve exhibits a crossover from velocity-weakening behavior to velocity-strengthening behavior (at $v_m$, here a few mm/s). This feature has been extensively discussed in \cite{BarSinai2012} and plays an important role below.

The initial conditions for the friction variables are selected so as to represent typical laboratory experiments $\tau^{el}(x,t\!=\!0)\!=\!300$kPa and $\phi(t\!=\!0)\!=\!1$s \cite{Ben-David2010a,Ben-David2010b}. The existence of an initial stress distribution $\tau^{el}(x,t\!=\!0)$ was shown to be a generic feature of frictional systems \cite{Ben-David2010a}, and -- as also shown below -- to affect the subsequent failure dynamics. Additional shear stresses are inhomogeneously applied to the system through moving its trailing edge at $x\!=\!0$ at a constant speed $v_d\!=\!10\mu$m/s, again typical to laboratory experiments \cite{Baumberger1999,Rubinstein2007,Ben-David2010a}. The resulting applied tangential force per unit thickness (out-of-plane) $f_d(t)$ is tracked.

\textit{Numerical results --- } We first characterize the phenomenology of the model through numerical simulations, a typical example of which is shown in Figs. \ref{fig:fig1}c-d. $f_d(t)$ is shown in Fig. \ref{fig:fig1}d to continuously curve (after a short quasi-linear increase) and to experience sharp, discrete-like, drops \cite{Rubinstein2007,Bouchbinder2011}.

To better characterize this behavior, we plot the corresponding spatiotemporal dynamics of $\tau(x,t)/\sigma$ in the color-map in Fig. \ref{fig:fig1}c (sharing the same time axis with Fig. \ref{fig:fig1}d). The continuous curving of $f_d(t)$ corresponds to the propagation of a {\em creep patch} that extends from $x\!=\!0$ into the interface and decelerates continuously (marked with a white arrow). When the creep patch reaches a certain size (marked by the horizontal dashed line), at $t\!\simeq\!16$s, it loses stability, and a much faster rupture front emerges and propagates until it arrests at $x\!\simeq\!10$cm. The rupture front propagation, responsible for the drop in $f_d(t)$ (marked by the black arrows in both panels), appears as a vertical line in the color-map because of the enormous variation in the time scales involved, though its velocity is finite (see below). A movie of the spatiotemporal dynamics is available at \cite{supp}.

When the rupture front arrests it leaves behind it an inhomogeneous stress distribution with a rather localized peak at the arrest location, which can be interpreted as the front tip. At the same time, another creep patch initiates and extends from the trailing edge until it loses stability at the {\em same size} as before and again a much faster rupture front propagates, collides with the previously arrested front tip and continues to propagate until it arrests deeper inside the interface (this time at $x\!\simeq\!14$cm). This process repeats itself almost periodically, though some heterogeneity appears (not discussed here).

\textit{Creep patches --- } A closer look at the creep patch is shown in Fig. \ref{fig:fig1}e, which presents a snapshot of the spatial distribution of the fields $v(x,t)$, $\tau(x,t)/\sigma$ and $A(x,t)$ at $t$ corresponding to the vertical dashed line in Fig. \ref{fig:fig1}c (prior to the instability). All fields relax to their spatially-homogeneous background values at the same point ($x\!\simeq\! 6$cm for that snapshot), which is the boundary between slipping and non-slipping regions, denoted by $x_{tip}$. To compute the creep patch velocity $c_{cr}\!\equiv\!\dot x_{tip}$, we assume that its dynamics are quasi-static and therefore neglect the inertial and viscous terms in Eq. (\ref{eq:udot}). We further replace $\tau^{el}$ by its fixed point to obtain $\bar{G} H \pa_{xx}u\!\simeq\!\mu_0 D A(\phi)/h$.

Transforming to a co-moving coordinate $\xi\!=\!x\!-\!c_{cr}t$ and estimating $\pa_x v\!\simeq\!v_d/x_{tip}$, the above relation yields
\begin{equation}
c_{cr} \simeq v_d \frac{\bar{G} H h }{\mu_0 D A(\phi_{tip})}\frac{1}{x_{tip}}\ , \label{eq:cofL}
\end{equation}
where $\phi_{tip}$ is an estimation of $\phi$ at the tip. This result shows that the creep patch propagation is directly driven by the loading as $c_{cr}$ is proportional to $v_d$ \cite{Bouchbinder2011}. Possibly related loading-rate dependent creep patches were observed in \cite{Ohnaka1999}. Moreover, Eq. (\ref{eq:cofL}) predicts that the creep patch decelerates as it extends, its propagation velocity being inversely proportional to its size, which is a property of the side-loading configuration. This prediction is verified in Fig. \ref{fig:fig1}f. Finally, we note that while $c_{cr}$ is significantly larger than the loading rate $v_d\!=\!10\mu$m/s -- in the cm/s range for our parameters here (cf. Fig. \ref{fig:fig1}f) -- it is still orders of magnitude slower than ``slow'' rupture \cite{Rubinstein2004, Nielsen2010, Ben-David2010a} and should not be confused with it.

\textit{Instability of creep patches (rupture nucleation) ---}
\begin{figure}
 \includegraphics[width=\columnwidth]{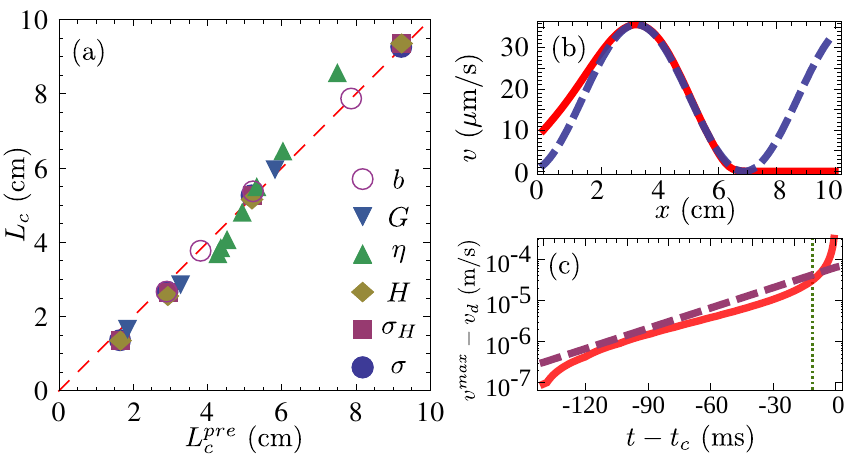}
 \caption{\textbf{Onset of instability.} (a) The nucleation size $L_c$ (measured in the simulation) vs. the prediction in Eq. (\ref{eq:Lc}). The parameters varied are shown in the legend (the dashed red line has a slope $1$ and goes through the origin). (b) Snapshot of $v(x,t)$ at the initial instability growth (solid red line, $t$ corresponds to the vertical line in (c)). $1\!+\!\cos{(2\pi x/L_c)}$, properly $x$-shifted and amplitude-scaled, is superimposed (dashed blue line). (c) The instability grows exponentially with time scale of 26ms. $v^{max}(t)$ is the instantaneous spatial maximum of $v(x,t)$ and $t_c$ is defined in Fig. \ref{fig:fig3}.}
\label{fig:fig2}
\end{figure}
Rapid slip nucleation (instability) at a critical size $L_c$ has been extensively discussed in the literature \cite{rice1983stability, Dieterich1992, Lapusta2000,Ohnaka1999,Ohnaka2000, Capozza2011} and is generically understood to result from a competition between frictional weakening and the variation of the effective bulk stiffness with the patch size. The advantage of the present framework is that it allows to analyze the instability very cleanly and carefully test the analytic predictions.

To analyze the stability of the creep patch we first note that its slip velocity is small and belongs to the weakening branch of the steady friction curve shown in Fig. \ref{fig:fig1}b. Therefore, we rewrite Eq. (\ref{eq:udot}) as $\bar{G} H \partial_{xx}u\!\simeq\!\tau\!\simeq\!\tau_{ss}(v)$, where $\tau_{ss}(v)$ is the velocity-weakening steady-state friction branch. We then introduce a displacement perturbation of the form $\delta u(x,t)\!=\!\delta u_0\,e^{ikx+\lambda t}$ in the above relation to obtain
\begin{equation}
 k^2 \bar{G}H\delta u\simeq \left|\partial \tau_{ss}/\partial v\right|\delta v\simeq
 \lambda\left|\partial \tau_{ss}/\partial v\right|\delta u\ ,\label{eq:pert2}
\end{equation}
resulting in an instability spectrum $\lambda\!\sim\!k^2$, in which larger $k$-vector modes grow faster. The spectrum is regularized by the intrinsic friction time scale, $\lambda\!\simeq\!v/D$, which yields for the most unstable mode $k_c\!=\!2\pi/L_c$ the following critical wavelength
\begin{equation}
 L_c \simeq 2\pi \sqrt{\frac{\bar{G} H D}{\left|\partial \tau_{ss}/\partial \log v\right|}} \label{eq:Lc} \ .
\end{equation}

The analysis above predicts that creep patches undergo a linear instability when $x_{tip}\!=\!L_c$, given in Eq. (\ref{eq:Lc}). This prediction is tested in detail in Fig. \ref{fig:fig2}. The dependence (and independence) of $L_c$ on various parameters in Eq. (\ref{eq:Lc}) is verified in Fig. \ref{fig:fig2}a. A snapshot of the velocity distribution during the initial growth of the instability is shown in Fig. \ref{fig:fig2}b. Superimposing $\cos{(2\pi x/L_c)}$ (i.e. the real part of $e^{ik_c x}$) on it yields excellent agreement (see Figure for details), which demonstrates that this is indeed a linear instability. Finally, our linear stability analysis predicts that $\lambda\!\simeq\!v_d/\!D\!\simeq\!(50$ms)$^{-1}$, where $v_d$ (the loading-rate) is the maximal slip velocity in the creep patch (cf. Fig. \ref{fig:fig1}e). Figure \ref{fig:fig2}c shows that the instability amplitude
initially grows exponentially with a typical time of $26$ms, in favorable agreement with the predictions. A movie of the instability is available at \cite{supp}.

To conclude the discussion of the instability we note that since $\left|\partial \tau_{ss}/\partial \log v\right|$ in the weakening regime is $v$-independent, $L_c$ in Eq. (\ref{eq:Lc}) is $v$-independent as well. Moreover, $L_c$ is independent of the stress state as is clearly demonstrated by the horizontal dashed line in Fig. \ref{fig:fig1}c (see below additional results concerning this point). The connection between Eq. (\ref{eq:Lc}) and available results in 2D is discussed in \cite{supp}.

\textit{Outcome of instability (rupture fronts) ---}
\begin{figure}
 \includegraphics[width=\columnwidth]{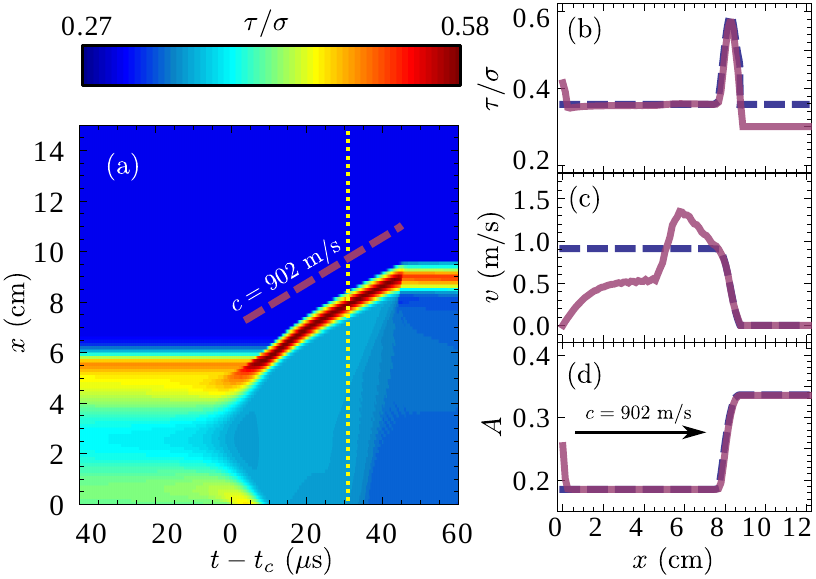}
 \caption{\textbf{Outcome of instability.} (a) High temporal resolution of the spatiotemporal evolution of $\tau(x,t)/\sigma$ during an instability (a movie is available at \cite{supp}). $t_c$ is defined as the zero of the time axis here (roughly at the onset of nonlinearity). (b)-(d) A snapshot of the fields distributions during rupture propagation at a time corresponding to the vertical dashed line in panel (a) (solid purple lines). The propagation velocity is $c\!=\!902$m/s. The dashed blue lines are described in the text.}
\label{fig:fig3}
\end{figure}
After an initial exponential growth, the instability enters the nonlinear regime, characterized by a steadily propagating rupture front that is excited for a few tens of $\mu$s and is accompanied by significant, much faster, slip (see Fig. \ref{fig:fig3}). What determines the rupture front properties?

In \cite{BarSinai2012} it was conjectured that transient rupture fronts propagating under spatially inhomogeneous stress conditions might be short-lived excitations of steady state rupture fronts propagating under homogeneous stress conditions. The latter exist only in the presence of a non-monotonic steady friction law (cf. Fig. \ref{fig:fig2}b) and span a continuous spectrum of propagation velocities with a finite minimal value \cite{BarSinai2012}. To test this idea, we choose a steady state front solution whose propagation velocity $c$ is the same as in Fig. \ref{fig:fig3} ($c\!=\!902$m/s, which is 32\% of the elastic wave-speed $c_s\!=\!\sqrt{\bar{G}/\rho}\!=\!2783$m/s) and which penetrates an interface of the same ``age'' (i.e. $\phi\!=\!17.4$s). When superimposing it on the transient front (solid purple lines in Figs. \ref{fig:fig3}b-d), we observe that all fields exhibit reasonable agreement, including the detailed distribution of $\tau(x,t)/\sigma$ and the typical slip velocity behind the front, lending support to our conjecture. Currently we cannot
theoretically predict the selection (i.e. why this particular $c$ was selected), which might be a ``soft selection'' due to the (weak) logarithmic velocity-strengthening.
\begin{figure}
 \includegraphics[width=\columnwidth]{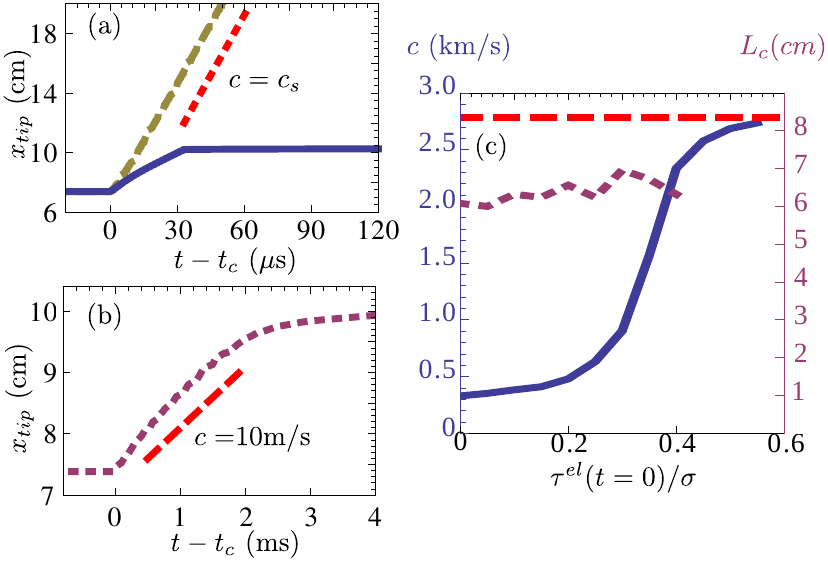}
 \caption{\textbf{Effect of velocity-strengthening and initial stress.} (a) The front location $x_{tip}$ during the first rupture event for logarithmic velocity-strengthening (solid blue line) and for purely velocity-weakening friction (dashed yellow line). The dotted red line corresponds to the elastic wave-speed. (b) $x_{tip}$ for linear strengthening. Note the dramatic change in the time scale as compared to panel (a). The dashed red line corresponds to $c\!=\!10$m/s. (c) $L_c$ (dotted purple line, right y-axis) and $c$ (solid blue line, left y-axis) vs. $\tau^{el}(t\!=\!0)/\sigma$. The dashed horizontal line is the elastic wave-speed.}
\label{fig:fig4}
\end{figure}

To further test this conjecture, and explore the role played by the velocity-strengthening branch in general, we study two variants of our model, one in which friction is purely velocity-weakening (cf. the dashed yellow line in Fig. \ref{fig:fig1}b) and one in which velocity-strengthening is linear in $v$ (cf. the dotted purple line in Fig. \ref{fig:fig1}b) \cite{Weeks1993,Heslot1994,Ditorro,Reches2010}. In the former case, rupture propagates at the elastic wave-speed $c_s$, penetrates much deeper into the interface and results in a much larger stress drop (see Fig. \ref{fig:fig4}a). In the latter case, rupture propagates at a much slower velocity $c\!\simeq\!10\hbox{m/s}\ll\!c_s$ (see Fig. \ref{fig:fig4}b), comparable to the smallest velocity member in the spectrum of steady state front solutions \cite{Bouchbinder2011, BarSinai2012}. We identify it as ``slow'' rupture \cite{Rubinstein2004, Nielsen2010}. These results clearly indicate that the existence and functional form of the velocity-strengthening branch significantly affect rupture dynamics. This might also explain why models that do not include velocity-strengthening friction typically feature only very fast rupture events \cite{Shi2010}.

Finally, we study the effect of the initial stress level on both the onset of instability and the resulting rupture (for logarithmic velocity-strengthening). Figure \ref{fig:fig4}c shows that the pre-stress $\tau^{el}(t\!=\!0)$ significantly affects the rupture propagation velocity (and hence the event's magnitude), while $L_c$ is almost unaffected (note that at $t\!=\!0$, $\tau^{el}(t\!=\!0)$ is balanced by $\pa_{xx} u$ in Eq. (\ref{eq:udot})). In a geophysical context, this result seems to agree with the statement that ``the size of an event is determined by the conditions on the fault segments the event is propagating into rather than by the nucleation process itself'' \cite{Lapusta2000}. In addition, we note that the variation of the rupture propagation velocity with the pre-stress level resembles the recent experimental results of \cite{Ben-David2010a} (cf. Fig. 3 therein).

\textit{Concluding remarks ---} In conclusion, we showed that creep patches extending at frictional interfaces undergo a linear instability at a critical nucleation size that is nearly independent of the stress state and the presence of velocity-strengthening friction. The post-instability nonlinear evolution results in rapid slip mediated by rupture fronts whose properties do depend on the stress state, the presence of velocity-strengthening friction and its functional form. In particular, the absence of velocity-strengthening friction facilitates large slip events that propagate at velocities approaching the elastic wave-speed and its presence gives rise to significantly smaller and slower slip events. Finally, we related transiently propagating rupture fronts to homogeneously-driven steady state fronts \cite{BarSinai2012} and showed that initial stresses systematically affect the rupture dynamics. These results may have significant implications for our understanding of interfacial failure and are currently extended to 2D.

\begin{acknowledgments}
EB acknowledges support of the James S. McDonnell Foundation, the Minerva Foundation with funding from the Federal German Ministry for Education and Research, the Harold Perlman Family Foundation and the William Z. and Eda Bess Novick Young Scientist Fund. EAB acknowledges support of the Erna and Jacob Michael visiting professorship funds at Weizmann Institute of Science.
\end{acknowledgments}


\end{document}